\documentclass[12pt,a4paper]{article}

\usepackage{hyperref}

\usepackage[top=2.0cm,bottom=2.0cm,left=2.0cm,right=2.0cm]{geometry}
\usepackage{latexsym,amsmath,amsfonts,amssymb,amsthm,amscd,mathrsfs}

\newcommand{\be}{\begin{equation}}           %
\newcommand{\ee}{\end{equation}}             %
\newcommand{\bea}{\begin{eqnarray}}          %
\newcommand{\eea}{\end{eqnarray}}            %
\newcommand{\ba}{\begin{aligned}}            %
\newcommand{\ea}{\end{aligned}}              %

\def\1{{\boldsymbol 1}}                     %
\def\cH{{\mathcal H}}                       %
\def\tr{\mathrm{tr}}                        %
\def\diag{\mathrm{diag}}                    %
\def\ri{{\rm i}}                            %
\def\C{\mathbb{C}}                          %
\def\N{\mathbb{N}}                          %
\def\T{\mathbb{T}}                          %
\def\UN{{\rm U}}                            %
\def\GL{{\rm GL}}                           %
\def\fH{\mathfrak{H}}                       %
\def\cF{{\mathcal F}}                       %
\def\fP{{\mathfrak{P}}}                     %
\def\reg{\mathrm{reg}}                      %
\def\red{\mathrm{red}}                      %
\def\cR{{\mathcal R}}                       %
\def\Ad{{\mathrm{Ad}}}                      %
\def\id{{\mathrm{id}}}                      %
\def\dt {\left.\frac{d}{dt}\right|_{t=0}}   %
\def\Herm{{\mathfrak{H}}}                   %
\def\fM{\mathfrak{M}}                       %
\def\u{\mathfrak{u}}                        %
\def\b{\mathfrak{b}}                        %
\def\B{\mathrm{B}}                          %
\def\gl{\mathfrak{gl}(n,\C)}                %
\def\cN{{\mathcal N}}                       %
\def\cinf{C^\infty}                         %
\def\half{\textstyle{\frac12}}              %
\def\TT{\mathbb{T}}                         %
\def\im{\mathrm{im}}                        %
\def\real{\mathrm{real}}                    %

\begin{document}

\begin{center}
{\Large\bf
On the bi-Hamiltonian structure of the trigonometric spin Ruijsenaars--Sutherland  hierarchy}
\end{center}

\medskip
\begin{center}
L.~Feh\'er${}^{a,b}$ and. I. Marshall${}^c$
\\

\bigskip
${}^a$Department of Theoretical Physics, University of Szeged\\
Tisza Lajos krt 84-86, H-6720 Szeged, Hungary\\
e-mail: lfeher@physx.u-szeged.hu

\medskip
${}^b$Department of Theoretical Physics, Wigner Research Centre for Physics\\
H-1525 Budapest, P.O.B.~49, Hungary\\

\medskip
${}^c$Faculty of Mathematics, Higher School of Economics, National Research University\\
Usacheva 6, Moscow, Russia\\
 e-mail: imarshall@hse.ru
\end{center}

\medskip
\begin{abstract}
We report on the the trigonometric spin Ruijsenaars--Sutherland hierarchy derived recently
by Poisson reduction of a bi-Hamiltonian hierarchy associated with
free geodesic motion on the Lie group $\UN(n)$.  In particular, we give a direct
proof of a previously stated result about the form of the second Poisson bracket in terms
of convenient variables.
\end{abstract}

\section{Introduction}

The classical integrable many-body models of Calogero--Moser--Sutherland  and Ruijsenaars--Schneider
as well as their extensions by internal degrees of freedom are in the focus of intense investigations  even today, many years
after their inception. See \cite{AO,CF,F1,F2} and references therein.
 One of the sources of these models is Hamiltonian reduction of
obviously integrable `free motion' on suitable higher dimensional phase spaces, among which
cotangent bundles and their Poisson--Lie analogues are the prime examples.  In this framework,
the emergence of the internal degrees of freedom, colloquially called `spin', originates
from the fact that symplectic reductions of cotangent bundles are in general not cotangent bundles,
but more complicated phase spaces.

We do not have a single, all encompassing framework for understanding integrable Hamiltonian systems, but there exist
several powerful approaches with large intersections of their ranges of applicability.
For example, the method of the classical $r$-matrix incorporates many famous systems, like Toda lattices, that can be
derived by Hamiltonian reduction, too, as reviewed in \cite{Per, RSTS}.
The $r$-matrix method and Hamiltonian reduction also have several links to the bi-Hamiltonian
approach initiated by Magri \cite{Ma}.

It was pointed out in  the recent paper \cite{F2} that one of the simplest finite-dimensional
integrable systems, the free geodesic motion on the unitary group $\UN(n)$, admits a natural bi-Hamiltonian
structure, and a suitable reduction of this free system gives rise to the so-called spin
Ruijsenaars--Sutherland hierarchy.
In this contribution, we overview the results of \cite{F2}, and give a new, direct proof of a statement
formulated in this reference without detailed proof.

\section{Bi-Hamiltonian hierarchy on $T^*\UN(n)$ and its reduction}

In this section we present a terse review of the results of \cite{F2}.

Our starting point is the manifold $T^* \UN(n)$, which we identify  with the set
\be
\fM:= \UN(n) \times \Herm(n):= \{ (g, L)\mid g\in \UN(n), L\in \Herm(n)\},
\label{H1}\ee
using right-trivialization. Here, the vector space of Hermitian matrices, $\Herm(n) = \ri \u(n)$,
serves  as the model of
the dual $\u(n)^*$ of the Lie algebra $\u(n)$.

Consider the \emph{real} Lie algebra $\gl$ endowed
with the non-degenerate bilinear form
\be
\left\langle X,Y \right\rangle:= \Im\tr(XY),
\quad\forall X,Y\in \gl.
\label{form}\ee
Then $\gl$ is the vector space direct sum of its isotropic Lie subalgebras
$\u(n)$ and $\b(n)$, where $\b(n)$ contains the upper triangular matrices with
real entries along the diagonal.
Consequently, we can decompose any $X\in \gl$ as
\be
X= X_{\u(n)} + X_{\b(n)}, \qquad X_{\u(n)}\in \u(n),\, X_{\b(n)}\in \b(n).
\label{Xdec}\ee
We also have another decomposition into
isotropic linear subspaces, $\gl = \u(n) + \Herm(n)$. Thus both $\b(n)$ and $\Herm(n)$ can serve as models
of $\u(n)^*$.

For any real function $F\in C^\infty(\fM)$,  introduce the derivatives
\be
D_1F, D_1' F \in C^\infty(\fM, \b(n)) \quad\hbox{and}\quad d_2 F \in C^\infty(\fM, \u(n))
\label{H2}\ee
by the relation
\be
\langle  D_1 F(g,L), X\rangle +
\langle  D_1' F(g,L), X'\rangle + \langle d_2 F(g,L), Y\rangle
= \dt F(e^{tX}g e^{t X'}, L+ t Y),
\label{H3}\ee
 for every $X,X'\in \u(n)$ and $Y\in \Herm(n)$.
The `free Hamiltonians' of our interest  are
\be
H_k(g,L):= \frac{1}{k} \tr(L^k),\quad
\forall k \in \N.
\label{Hk}\ee
These feature in the `free bi-Hamiltonian hierarchy' on $\fM$, which is given by the next theorem.

\medskip\noindent
{\bf Theorem 1 \cite{F2}.} \emph{The following formulae define two compatible Poisson brackets on $\fM$:
\be\{F, H\}_1(g,L) =
\left\langle D_1F , d_2H \right\rangle - \left\langle D_1H, d_2F \right\rangle
+ \left\langle L,  [d_2F,  d_2H] \right\rangle,
\label{H4}\ee
and
\bea
&&\{F, H\}_2(g,L) =  \left\langle D_1F , Ld_2H\right\rangle
-\left\langle D_1H, L d_2F \right\rangle
\nonumber \\
&&\qquad \qquad  + 2\left\langle L d_2F, \left(L d_2H \right)_{\u(n)}\right\rangle
-\frac{1}{2} \left\langle  D'_1F , g^{-1} (D_1H) g \right\rangle,
\label{H5}\eea
where the derivatives are taken at $(g,L)$ and (\ref{Xdec}) is applied.
The Hamiltonians $H_k$ satisfy
\be
\{ F, H_k\}_2 = \{ F, H_{k+1}\}_1,
\qquad
\forall F\in C^\infty(\fM),
\label{biH}\ee
and $\{H_k, H_\ell\}_1 = \{ H_k, H_\ell\}_2$ =0 for every $k, \ell\in \N$.
 The bi-Hamiltonian flow
of the systems $\left(\fM,\{\ ,\ \}_2, H_k\right)$ and  $\left(\fM,\{\ ,\ \}_1, H_{k+1}\right)$ is given by
$(g(t), L(t)) = \left( \exp( \ri t L(0)^k) g(0), L(0)\right)$.
}

\medskip
The first Poisson bracket is the canonical one carried by the cotangent bundle of $\UN(n)$, while
the second one arises from the Heisenberg double \cite{STS} of the Poisson--Lie group  $\UN(n)$.
The latter point is explained in \cite{F2},  where it is also noted that the Lie derivative of the
Poisson tensor of  $\{\ ,\ \}_2$
along
the infinitesimal generator of the flow  $(g(t), L(t)) = (g(0), L(0) + t \1_n)$ is the Poisson tensor of $\{\ ,\ \}_1$.
 This implies \cite{Sm} compatibility, and the rest of the statements
is readily checked as well.

The fact that the flow generated by the Hamiltonian $H_1$ on the Heisenberg double of $\UN(n)$ projects to free motion on $\UN(n)$
was pointed out long time ago  by S. Zakrzewski \cite{Z}, which served as one of the motivations behind Theorem 1.

The `conjugation action' of $\UN(n)$ on $\fM$ associates with every $\eta\in \UN(n)$
the diffeomorphism $A_\eta$ of $\fM$ that operates according to
\be
A_\eta(g,L) := (\eta g \eta^{-1}, \eta L \eta^{-1}).
\label{Act1}\ee
 A key property of the Poisson brackets on $\fM$ is that they can be restricted to the set of invariant functions with respect
 to this action, denoted $C^\infty(\fM)^{\UN(n)}$.
 This means that if $F,H\in C^\infty(\fM)^{\UN(n)}$, then the same holds
 for their Poisson brackets $\{F,H\}_i$ for $i=1,2$.
 Because the Hamiltonians $H_k$ are also invariant, we can restrict the `free hierarchy' to $\UN(n)$-invariant
 observables. This procedure,  called Poisson reduction \cite{RSTS}, is an algebraic formulation of
 projection onto the quotient space $\fM/\UN(n)$.

 Any smooth function on $\fM$  can be recovered from its restriction to the dense open
 submanifold $\fM_\reg \subset \fM$, which contains the points $(g,L)$ with $g$ having
 distinct eigenvalues. Moreover, $F\in C^\infty(\fM_\reg)^{\UN(n)}$ is uniquely determined by its restriction
 $f$ on the manifold $\T^n_\reg \times \fH(n)$, where $\T^n_\reg$ is the set of regular elements
 in the standard maximal torus of $\UN(n)$. In fact, restriction engenders a one-to-one
 correspondence
 \be
 C^\infty(\fM_\reg)^{\UN(n)} \longleftrightarrow C^\infty(\T^n_\reg\times \Herm(n))^{\cN(n)},
\label{corr}\ee
where $\cN(n)$ is the normalizer of $\T^n$ in $\UN(n)$, whose action preserves $\T^n_\reg\times \Herm(n)$.
Note that $\cN(n)$ is the semi-direct product of the permutation group $S_n$,
naturally embedded into $\UN(n)$, with $\T^n$.
By taking advantage of the correspondence (\ref{corr}),  one can encode the Poisson brackets on $C^\infty(\fM_\reg)^{\UN(n)}$
by two compatible Poisson brackets $\{\ ,\ \}_i^\red$ on $C^\infty(\T^n_\reg\times \Herm(n))^{\cN(n)}$.
The main result of \cite{F2} is the formula of these reduced Poisson brackets.

For  $f\in C^\infty(\T^n_\reg \times \Herm(n))$,  the
$\b(n)_0$-valued derivative $D_1 f$ and the $\u(n)$-valued derivative $d_2 f$ are defined
by the equality
\be
\langle  D_1 f(Q,L), X\rangle
+ \langle d_2 f(Q,L), Y\rangle
= \dt f(e^{tX}Q , L+ t Y),
\label{Dfdef}\ee
 for every $X\in \u(n)_0$ and $Y\in \Herm(n)$, where $\b(n)_0$ and $\u(n)_0$ denote
 the subalgebras of diagonal matrices in $\b(n)$ and $\u(n)$, respectively.
Decompose $\gl$ as the vector space direct sum of subalgebras
 \be
 \gl= \gl_+ + \gl_0 + \gl_-,
 \label{N7}\ee
defined by means of the principal gradation.
Accordingly, we can decompose any $X\in \gl$ as $X= X_+ + X_0 + X_-$, where $X_0$ is diagonal and $X_+$ is strictly upper-triangular.
 Then, for $Q\in \T_\reg^n$,  introduce $\cR(Q)\in \mathrm{End}(\gl)$ by setting it equal to zero on $\gl_0$ and
 defining it otherwise as
 \be
 {\cR(Q)\vert}_{\gl_+ + \gl_-} := \frac{1}{2}(\Ad_Q + \id) \circ \left({(\Ad_Q - \id)\vert}_{\gl_+ + \gl_-}\right)^{-1},
 \label{N8}\ee
 where $\Ad_Q(X)= Q X Q^{-1}$ for all $X\in \gl$. The definition makes sense because of the regularity of $Q$.
 Note that $\langle \cR(Q)X, Y\rangle = - \langle X, \cR(Q) Y\rangle$, and
introduce the notation
 \be
 [X,Y]_{\cR(Q)} := [ \cR(Q) X, Y] + [X, \cR (Q)Y],
 \quad\forall X,Y\in \gl.
 \label{N9}\ee

\medskip
\noindent
{\bf Theorem 2 \cite{F2}.}
\emph{For $f, h\in C^\infty(\T_\reg^n \times \Herm(n))^{\cN(n)}$, the reduced Poisson brackets
 have the form
\be
\{ f, h\}_1^\red(Q,L) =
\langle D_1 f, d_2 h\rangle - \langle D_1 h, d_2 f\rangle +\langle L, [ d_2 f, d_2 h]_{\cR(Q)}\rangle,
\label{N13}\ee
and
\be
\{ f, h\}_2^\red(Q,L) =
\langle D_1 f, L d_2 h\rangle - \langle D_1 h, L d_2 f\rangle + 2\langle L d_2 f, \cR(Q) (L d_2 h)\rangle,
\label{N14}\ee
where the derivatives are evaluated at  $(Q,L)$, and the notations (\ref{N8}), (\ref{N9}) are applied.}
\medskip

The reduced system that descends from the free hierarchy generated the Hamiltonians $H_k$ (\ref{Hk}) is
called \emph{spin Ruijsenaars--Sutherland hierarchy}.
The reason for this terminology will become clear in the next section.
For the reduced equations of motion and remarks on their integrability, see \cite{F2}.

\section{Useful changes of variables}

In the first subsection we introduce new variables that behave as
 canonically conjugate pairs and `spin variables'  with respect to the second Poisson bracket,
 and allow us to interpret $\tr(L)$ as a spin Ruijsenaars Hamiltonian.
 These new variables go back to the papers \cite{F1,F2}.
 In the second subsection we describe another, in this case well-known \cite{FP,LX}, set of new variables,
 which convert the first Poisson bracket into that of canonical pairs and (other kind of) spin variables,
 and lead to the interpretation of $\tr(L^2)$ as a spin Sutherland Hamiltonian.

\subsection{Interpretation as spin Ruijsenaars model}

We now discuss the change of variables that the underlie the interpretation
of the reduced free system as a spin Ruijsenaars model. For this purpose, we focus on the second Poisson bracket (\ref{N14}),
and restrict ourselves to the open submanifold
\be
\T^n_\reg \times \fP(n) \subset  \T^n_\reg \times \Herm(n),
\ee
where $\fP(n)$ denotes the set of positive definite Hermitian matrices.
It is a standard fact of linear algebra that any $L\in \fP(n)$ can be uniquely written in the form
\be
L= bb^\dagger\,\,\hbox{with}\,\, b\in \B(n),
\label{Lfact}\ee
and $b\in \B(n)$ can be decomposed as
\be
b= e^p b_+  \,\,\hbox{with}\,\,
p\in \b(n)_0,\,\,b_+\in \B(n)_+,
\label{bfact}\ee
where $\B(n)_+$ is the group of
upper triangular matrices with unit diagonal.
We  define
\be
\lambda:= b_+^{-1} Q^{-1} b_+ Q,
\label{lambda}\ee
and obtain the  change of variables
\be
\T^n_\reg \times \fP(n)\ni (Q, L) \longleftrightarrow
(Q,p, \lambda) \in \T^n_\reg \times \b(n)_0 \times \B(n)_+.
\label{F1}\ee
A grade by grade inspection of the defining relation (\ref{lambda}) shows that this
is a diffeomorphism between the respective spaces. Thus every function $f(Q,L)$ corresponds to
a unique function $\cF(Q,p,\lambda)$.  The diffeomorphism  (\ref{F1}) induces an action of $\cN(n)$ on
$\T^n_\reg \times \b(n)_0 \times \B(n)_+$, and we are interested in the invariant functions.
The action of the subgroup $\T^n < \cN(n)$ is especially simple, it is given by
\be
(Q,p, \lambda) \mapsto (Q,p, \tau \lambda \tau^{-1}), \quad \forall \tau\in \T^n,
\ee
since this corresponds to $(Q,L) \mapsto (Q, \tau L \tau^{-1})$.

For any $\cF \in C^\infty( \T^n_\reg \times \b(n)_0 \times \B(n)_+)$, we define the derivatives
$D_Q\cF \in \b(n)_0$, $d_p \cF = \u(n)_0$ and
$D_\lambda\cF$, $D_\lambda'\cF \in \u(n)_\perp$ by
 \be
 \dt \cF(e^{tX_0} Q,p + t Y_0, e^{tX_+}\lambda  e^{t Y_+})=
 \langle D_Q\cF, X_0\rangle + \langle d_p \cF, Y_0\rangle +
\langle  D_\lambda \cF, X_+\rangle  + \langle D'_\lambda \cF , Y_+\rangle.
\label{F3}\ee
Here, $X_0\in \u(n)_0$, $Y_0\in \b(n)_0$
and
 $X_+, Y_+\in \b(n)_+$ are arbitrary, the argument $(Q,p,\lambda)$ is suppressed on  the right hand side, and
 $\u(n)_\perp$ denotes the off-diagonal linear subspace of $\u(n)$.

The next proposition was stated previously without elaborating its proof.

\medskip
\noindent
{\bf Proposition 3 \cite{F2}.} \emph{Consider the functions $\cF,
\cH \in C^\infty(\T^n_\reg \times \b(n)_0 \times \B(n)_+)^{\cN(n)}$
that are related to $f,h\in C^\infty(\T^n_\reg \times \fP(n))^{\cN(n)}$ according to
\be
\cF(Q,p,\lambda) = f(Q,L),\,\, \cH(Q,p,\lambda) = h(Q,L)\,\,
\hbox{with}\,\,
L = e^p b_+ b_+^\dagger e^p, \,\, \lambda:= b_+^{-1} Q^{-1} b_+ Q.
\ee
In terms of the variables $(Q,p,\lambda)$, the second Poisson bracket (\ref{N14}) takes the form
\be
2 \{\cF, \cH\}^\red_2(Q,p,\lambda) = \langle D_Q \cF, d_p \cH \rangle -
\langle D_Q \cH, d_p \cF \rangle
+ \langle  D'_\lambda \cF, \lambda^{-1} (D_\lambda \cH) \lambda \rangle,
\label{F2}\ee
where the derivatives are evaluated at $(Q,p,\lambda)$.}

\begin{proof}
Recall that $(Q,L)$, $(Q,b)$ and $(Q,p,\lambda)$ are alternative sets of variables. In particular,
we have the  invertible correspondences:
\be
(Q,L) \leftrightarrow  (Q,b) \leftrightarrow  (Q,p,\lambda) \quad\hbox{with}\quad \ L= bb^\dagger,\ e^p:=b_{\diag},\
\lambda := b^{-1}Q^{-1}bQ.
\ee
Here, we suppressed that $\lambda$ does not depend on $p$.
Any tangent vector at a fixed $(Q,b)$ can be represented as the velocity vector at $t=0$ of a curve of the form
\be
(Q(t), b(t)) =  (e^{t\xi}Q,\  be^{t\beta}), \,\, \hbox{with some}\,\, \xi\in \u(n)_0,\, \beta=(\beta_0 + \beta_+)\in \b(n).
\ee
In terms of the alternative variables, the corresponding curves are easily seen to satisfy
\be
\ba
&L(t) =L + tb(\beta+\beta^\dagger)b^\dagger +\mathrm{o}(t),\\
&\lambda(t) = \lambda\exp\bigl(t\bigl[\xi-Q^{-1}b^{-1}\xi bQ + Q^{-1}\beta Q - \lambda^{-1}\beta\lambda\bigr]
+\mathrm{o}(t)\bigr),\\
&p(t) = p + t \beta_0 + \mathrm{o}(t).
\ea
\ee
Of course, the curve that appears in the exponent after $\lambda$ lies in $\b(n)_+$.
Let us now consider a function on our space, which is either expressed as
$(Q,L)\mapsto  f(Q,L)$, or equivalently as $(Q,p,\lambda)\mapsto \cF(Q,p,\lambda)$.
By the definition of derivatives, we obtain the equality
\be
\ba
&\dt  f\bigl(Qe^{t\xi}, L + tb(\beta+\beta^\dagger)b^\dagger  + \mathrm{o}(t)\bigr)
=\\
&\qquad\quad\dt \cF\bigl( Qe^{t\xi}, p+t\beta_{0},
\lambda\exp(t[\xi-Q^{-1}b^{-1}\xi bQ + Q^{-1}\beta Q - \lambda^{-1}\beta\lambda]+\mathrm{o}(t))\bigr).
\ea
\ee
This generates the following relations between the derivatives of $f$ and $\cF$:
\be\label{derivcomp}
\ba
&\langle 2b^\dagger d_2f b - d_p \cF - QD'_\lambda \cF Q^{-1} +
(\lambda D'_\lambda \cF \lambda^{-1})_{\u(n)} \,,\,\beta\rangle
\\
&\qquad+  \langle D_1 f - D_Q\cF - D'_\lambda \cF + bQD'_\lambda \cF Q^{-1}b^{-1}\, , \, \xi\rangle= 0,
 \quad \forall\xi\in \u(n)_0,\,\forall\beta\in \b(n).
\ea
\ee
The derivatives of $f$ and $\cF$ are taken at $(Q,L)$ and at $(Q,p,\lambda)$, respectively,
according to
(\ref{Dfdef}) and (\ref{F3}).
We have $\langle D'_\lambda \cF, \xi \rangle =0$,
and the conventions  $D_\lambda' \cF,\, D_\lambda \cF \in \u(n)_\perp$ imply
\be
(\lambda D'_\lambda \cF \lambda^{-1})_{\u(n)} = D_\lambda \cF +  (\lambda D'_\lambda \cF \lambda^{-1})_{\im-\diag}.
\label{myders}\ee
The matrix $X_{\im-\diag}$ is obtained from the matrix $X$ by setting to zero the off-diagonal entries and
the real parts of the diagonal entries of $X$, and (\ref{Xdec}) is used.

From the first term in (\ref{derivcomp}) (the one involving arbitrary $\beta$), we must have
\be
A:=2b^\dagger d_2 f b - d_pF - QD'_\lambda \cF Q^{-1} + (\lambda D'_\lambda \cF \lambda^{-1})_{\u(n)}\in \b(n).
\ee
But the formula of $A$ shows that  $A\in\u(n)$, and thence $A=0$. It is convenient to rewrite
\be
2b^\dagger d_2 f b = QD'_\lambda\cF Q^{-1} - \lambda D'_\lambda \cF\lambda^{-1} + \bigl[ d_p\cF +
\lambda D_\lambda' \cF\lambda^{-1} - (\lambda D'_\lambda \cF \lambda^{-1})_{\u(n)}\bigr],
\ee
and, conjugating by $b$ and using $b\lambda= Q^{-1}bQ$, we get
\be\label{LdF}
\ba
2L\,d_2 f &= b QD'_\lambda \cF Q^{-1}b^{-1} - b \lambda D'_\lambda \cF\lambda^{-1}b^{-1} +
\Ad_b[d_p \cF+ \Ad_\lambda D_\lambda'\cF - (\lambda D'_\lambda \cF \lambda^{-1})_{\u(n)}]\\
&=
(\Ad_Q-\id)\Ad_{Q^{-1}bQ}D'_\lambda \cF + \Ad_b[d_p\cF+ \Ad_\lambda D_\lambda'\cF -
(\lambda D'_\lambda \cF \lambda^{-1})_{\u(n)}],
\ea
\ee
from which it is easy to obtain
\be\label{RLdF}
\ba
2 \cR(Q)(Ld_2 f) &=\frac12 (\Ad_Q+\id)\Ad_{Q^{-1}bQ}D'_\lambda \cF - (bQD'_\lambda \cF Q^{-1}b^{-1})_{\diag}\\
&\qquad+ \cR(Q)\bigl(\Ad_b[d_p\cF+ \Ad_\lambda D_\lambda'\cF -
(\lambda D'_\lambda \cF \lambda^{-1})_{\u(n}]  \bigr).
\ea
\ee
Of course, we could have written everywhere
$\Ad_\lambda D_\lambda'\cF - (\lambda D'_\lambda \cF \lambda^{-1})_{\u(n)} \equiv (\Ad_\lambda D_\lambda'\cF )_{\b(n)}$.
Note also that $\Ad_m$ denotes conjugation by $m$ for any $m\in \GL(n,\C)$.

A glance at the last equation (\ref{RLdF}) shows that the expression
in the second line belongs to $\b(n)_+$, and this is crucial for the computation of
$\langle Ld_2f, \cR(Q)(Ld_2h)\rangle$ (cf. (\ref{N14})):
\be\label{LdFRLdH}
\ba
&4\langle Ld_2 f\,,\, \cR(Q)(Ld_2 h)\rangle=\\
&\quad
\langle
(Ad_Q-\id)\Ad_{Q^{-1}bQ}D'_\lambda \cF + \Ad_b[d_p\cF+ \Ad_\lambda D_\lambda'\cF -
(\lambda D'_\lambda \cF \lambda^{-1})_{\u(n)}] \,, - (\Ad_{bQ}D'_\lambda \cH)_{\diag} \\
&\qquad
+\half (\Ad_Q+\id)Ad_{Q^{-1}bQ}D'_\lambda \cH
+ \cR(Q)\bigl(\Ad_b[d_p \cH+ \Ad_\lambda D_\lambda'\cH - (\lambda D'_\lambda \cH \lambda^{-1})_{\u(n)}]\bigr)
\rangle\\
&=
\half \langle
\Ad_{bQ}D'_\lambda \cF , Ad_{Q^{-1}bQ}D'_\lambda \cH
\rangle \\
&\qquad+\half \langle
d_p\cF+ \Ad_\lambda D_\lambda'\cF - (\lambda D'_\lambda \cF \lambda^{-1})_{\u(n)}  \, , \, \Ad_Q D_\lambda' \cH +
\Ad_\lambda D'_\lambda \cH - 2(\Ad_{bQ}D'_\lambda \cH)_{\diag}
\rangle\\
&\qquad - (\cF\leftrightarrow \cH)
\\
&=
\half\langle \Ad_QD'_\lambda \cF , \Ad_\lambda D'_\lambda \cH\rangle
+\half \langle d_p\cF , \Ad_\lambda D'_\lambda \cH - 2 bQD'_\lambda \cH Q^{-1}b^{-1}\rangle  +
\half\langle \Ad_\lambda D_\lambda'\cF ,\Ad_Q D_\lambda'\cH\rangle\\
&\qquad
+
\langle  (\lambda D'_\lambda \cF \lambda^{-1})_{\u(n)} -
\Ad_\lambda D_\lambda'\cF , (\Ad_{bQ}D'_\lambda \cH )_{\diag}\rangle
- \half\langle (\lambda D'_\lambda \cF \lambda^{-1})_{\u(n)} , \Ad_\lambda D_\lambda'\cH\rangle \\
&\qquad - (\cF\leftrightarrow \cH).
\ea
\ee
Notice that the first and fourth terms  in the first line after the last equality sign add up to
\be
\half\langle \Ad_QD'_\lambda \cF , \Ad_\lambda D'_\lambda \cH\rangle + \half\langle Ad_\lambda D_\lambda'\cF,Ad_QD_\lambda'\cH\rangle,
\ee
and this is symmetric with respect to exchange of $\cF$ and $\cH$; thereby it cancels.
Notice also that the first expression in the second line simplifies as follows:
\be
\ba
 &\langle  (\lambda D'_\lambda \cF \lambda^{-1})_{\u(n)} - \Ad_\lambda D_\lambda'\cF , (\Ad_{bQ}D'_\lambda \cH )_{\diag}\rangle \\
 &\qquad \qquad = \langle  (\lambda D'_\lambda \cF \lambda^{-1})_{\u(n)} -
 \Ad_\lambda D_\lambda'\cF , (\Ad_{bQ}D'_\lambda \cH )_{\im-\diag}\rangle\\
 &\qquad \qquad  =- \langle   \Ad_\lambda D_\lambda'\cF , (\Ad_{bQ}D'_\lambda \cH )_{\im-\diag}\rangle,
\ea
\ee
 which will be shortly shown to vanish.
To summarize, we obtained
\be
\ba
 &4\langle Ld_2 f\,,\, \cR(Q)(Ld_2 h)\rangle=
 - \half \langle \Ad_\lambda D_\lambda'\cF, d_p \cH +2(\Ad_{bQ}D'_\lambda \cH )_{\im-\diag}\rangle
- \langle d_p \cF, \Ad_{bQ} D'_\lambda \cH \rangle
\\
&\qquad\qquad -  \half\langle (\lambda D'_\lambda \cF \lambda^{-1})_{\u(n)} ,
\Ad_\lambda D_\lambda'\cH\rangle - (\cF\leftrightarrow \cH ).
\ea
\label{good}\ee

Next, we may look at the other terms, and return to the $\xi$-term of (\ref{derivcomp}). This gives
\be
 D_1 f = D_QF - (\Ad_{bQ}D'_\lambda \cF)_{\real- \diag},
\ee
which, together with (\ref{LdF})---discarding the term in the range of $(\Ad_Q-\id)$ as this
 is in the annihilator of $\b(n)_0$---gives us
\be\label{DFLdH}
\ba
2\langle D_1f\,,\, Ld_2 h\rangle&=
\langle D_Q\cF - (\Ad_{bQ}D'_\lambda \cF)_{\real-\diag}\,,\,
\Ad_b[d_p\cH+ \Ad_\lambda D_\lambda'\cH -  (\lambda D_\lambda'\cH \lambda^{-1})_{\u(n)}] \rangle\\
&=
\langle
D_Q\cF - (\Ad_{bQ}D'_\lambda \cF)_{\real-\diag} \,,\, d_p\cH\rangle= \langle D_Q\cF - \Ad_{bQ}D'_\lambda \cF \,,\, d_p\cH\rangle.
\ea
\ee
Putting together now (\ref{good}) and (\ref{DFLdH}), the second term at the very end of (\ref{DFLdH})
cancels, and we arrive at
\be\label{newpb1}
\ba
&2\{f, h\}_2^\red (Q,L) = 2\langle D_1 f, Ld_2 h\rangle -
 2\langle Ld_2 f, D_1 h\rangle + 4\langle Ld_2 f, \cR(Q)(Ld_2 h)\rangle\\
\qquad &=
\langle
D_Q\cF \,,\, d_p\cH \rangle + \half \langle \Ad_\lambda D_\lambda'\cF,  (\lambda D'_\lambda \cH \lambda^{-1})_{\u(n)} \rangle
- \half \langle \Ad_\lambda D_\lambda'\cF, \eta_{\cH}\rangle
 -(\cF\leftrightarrow \cH),
\ea
\ee
where $\u(n)_0\owns \eta_{\cH}: = d_p\cH + 2(\Ad_{bQ}D_\lambda'\cH )_{\im-\diag}$ represents the diagonal-imaginary
entities from the previous formulae.
As explained below, for invariant functions $\cF$ and $\cH$,  the term containing $\eta_\cH$ vanishes,  and
we also have
\be
\langle \Ad_\lambda D_\lambda'\cF,  (\lambda D'_\lambda \cH \lambda^{-1})_{\u(n)} \rangle =
\langle \Ad_\lambda D_\lambda'\cF,   D_\lambda \cH  + (\lambda D'_\lambda \cH \lambda^{-1})_{\im-\diag}\rangle
=\langle \Ad_\lambda D_\lambda'\cF,   D_\lambda \cH  \rangle,
\ee
where we used (\ref{myders}) and the property (\ref{van}).

By the above, the claim of the proposition follows from (\ref{newpb1}) if we can verify that
for any $\cF\in\cinf(\TT^n_\mathrm{reg}\times\b(n)_0\times \B(n)_+)^{\TT^n}$ we have
\be
\langle X , \lambda D'_\lambda \cF\lambda^{-1}\rangle =0,
\quad  \forall X\in \u(n)_0.
\label{van}\ee
In order to justify this, we remark that
\be
\langle X,\lambda D'_\lambda \cF\lambda^{-1}\rangle =
\langle \lambda^{-1}X\lambda -X ,D'_\lambda F\rangle.
\ee
Since $\lambda^{-1}X\lambda-X\in\b(n)_+$, we may rewrite this as
\be
\langle X,\lambda D'_\lambda \cF\lambda^{-1}\rangle = \dt \cF\bigl(Q,p,\lambda\exp(t[\lambda^{-1}X\lambda-X])\bigr)
=\dt \cF\bigl(Q,p,e^{tX}\lambda e^{-tX}\bigr).
\label{last}\ee
In the last step we used that $\dt \lambda\exp(t[\lambda^{-1}X\lambda-X]) = [X,\lambda]$.
We see from  (\ref{last}) that (\ref{van}) follows from the $\T^n$-invariance of $\cF$,
and hence  the proof is complete.
\end{proof}

Regarding the interpretation of Proposition 3,
it is worth pointing out that one may view the restriction to $\cN(n)$-invariant functions on
$\T^n_\reg\times \b(n)_0 \times \B(n)_+$  as the result of a  two step process.
The first step consists in Hamiltonian reduction  of
$\T^n_\reg\times \b(n)_0 \times \B(n)$ by the
normal subgroup $\T^n$.
The formula (\ref{F2}) defines a Poisson bracket already on the $\T^n$-invariant functions.
In fact, its last term can be identified as the  result of reduction of the multiplicative Poisson
bracket on $\B(n)$ by the conjugation action of $\T^n$, at the zero value of the pertinent moment map.
In other words, the last term of (\ref{F2}) corresponds to the Poisson space  $\B(n)//_0 \T^n$.
(Cf. Theorem 4.3 in \cite{F1}.)
The second step consists in taking quotient by $S_n=\cN(n)/\T^n$.

When expressed in the variables $(Q,p,\lambda)$,
 the Hamiltonian $\tr(L)= \tr(bb^\dagger)= \tr(e^{2p} b_+ b_+^\dagger)$ can be written as
 \be
 \tr(L) = \sum_{i=1}^n e^{2p_i} V_i(Q,\lambda) \quad\hbox{with}\quad
 V_i(Q,\lambda)=\left(b_+(Q,\lambda) b_+(Q,\lambda)^\dagger\right)_{ii},
 \label{F4}\ee
where $\lambda$  is a `spin' variable,
and $b_+(Q,\lambda)$ denotes the solution of the equation (\ref{lambda}) for $b_+$.
An explicit formula of $b_+(Q,\lambda)$ can be extracted from Section 5.2 in \cite{F1}.
Comparison of (\ref{F4}) with the light-cone Hamiltonians of the standard RS model \cite{RS} justifies
calling this a   \emph{spin Ruijsenaars type Hamiltonian}.
A further justification is that restriction of the system to a one-point symplectic leaf
in $\B(n)//_0 \T^n$ yields the spinless trigonometric RS model \cite{FK1}.

\subsection{Interpretation as spin Sutherland model}

Concentrating on the first Poisson bracket (\ref{N13}),
we present another set of useful variables
\be
(Q, p, \phi) \in \T^n_\reg \times \Herm(n)_0 \times \Herm(n)_\perp,
\label{S4}\ee
where the subscripts $0$ and $\perp$ refer to diagonal matrices and
off-diagonal matrices, respectively.
The relevant change of variables is encoded by the diffeomorphism
\be
\gamma:  \T^n_\reg \times \Herm(n)_0 \times \Herm(n)_\perp \to \T^n_\reg \times \Herm(n)
\label{S5}\ee
operating according to
\be
\gamma: (Q, p, \phi) \mapsto (Q,L(Q,p, \phi)) \quad\hbox{with}\quad
L(Q, p,\phi)= p - (\cR(Q) + \frac{1}{2}\id )(\phi).
\label{S6}\ee
We now express the functions $f, h\in C^\infty(\T_\reg^n \times \Herm(n))^{\cN(n)}$ in the form
\be
f \circ \gamma = \cF,\,\, h\circ \gamma = \cH,
\quad
\cF, \cH \in C^\infty(\T_\reg^n \times \Herm(n)_0 \times \Herm(n)_\perp)^{\cN(n)},
\label{S7}\ee
where $\cN(n)$ acts in the natural manner inherited from the conjugation action.
The Poisson bracket $\{\ ,\ \}_1^\red$ on $C^\infty(\T_\reg^n \times \Herm(n)_0 \times \Herm(n)_\perp)^{\cN(n)}$
is defined by the formula
\be
\{ \cF, \cH\}_1^\red \equiv \{ \cF\circ \gamma^{-1}, \cH \circ \gamma^{-1}\}_1^\red \circ \gamma,
\label{S8}\ee
where (\ref{S6}) is used and the right-hand side refers to the Poisson bracket (\ref{N13}).

For any $\cF \in C^\infty(\T_\reg^n \times \Herm(n)_0 \times \Herm(n)_\perp)$, we have
the derivatives
\be
D_Q \cF(Q,p,\phi)\in \b(n)_0,
\quad
d_p \cF(Q,p,\phi)\in \u(n)_0,
\quad
d_\phi \cF(Q,p,\phi)\in \u(n)_\perp,
\label{S9}\ee
defined by
\be
\langle  D_Q \cF(Q,p,\phi), X\rangle +
\langle  d_p \cF(Q,p,\phi), Y_0\rangle + \langle d_\phi \cF(Q,p,\phi), Y_\perp\rangle
= \dt \cF(e^{tX}Q, p + t Y_0, \phi + t Y_\perp),
\label{S10}\ee
 for every $X\in \u(n)_0$ and $Y= (Y_0 + Y_\perp)\in \Herm(n)$.

\medskip\noindent
{\bf Proposition 4 \cite{FP,LX}.}
\emph{In terms of the variables  $(Q,p,\phi)$ defined by (\ref{S6}), the reduced first Poisson bracket (\ref{N13})
has the following form:
\be
\{ \cF, \cH\}_1^\red(Q,p,\phi) = \langle D_Q\cF, d_p \cH\rangle - \langle D_Q \cH, d_p \cF\rangle
+ \langle \phi, [ d_\phi \cF, d_\phi \cH]\rangle.
\label{S11}\ee
Here, $\cF, \cH\in C^\infty(\T_\reg^n \times \Herm(n)_0 \times \Herm(n)_\perp)^{\cN(n)}$ and
the derivatives are taken at $(Q,p,\phi)$.}

\medskip
The change of variables $(Q,L) \leftrightarrow (Q,p,\phi)$ appeared in the construction of spin Sutherland
models via the method of Li and Xu \cite{LX}, whose relation to Hamiltonian reduction of  free motion
on Lie groups  was clarified in \cite{FP}. The proof of Proposition 4 can be extracted
from these references.  One can also prove it by direct calculation, which is
much simpler than the one required for the proof of Proposition 3.

The reduced Hamiltonians $\cH_k^\red$ arising from those in (\ref{Hk}) can be written
in terms of the variables $(Q,p,\phi)$ as
\be
\cH_k^\red(Q,p,\phi) = \frac{1}{k} \tr( L(Q,p,\phi)^k).
\label{S21}\ee
For $k=2$, with
$Q = \exp\left(\diag(\ri q_1,\dots, \ri q_n)\right)$, and  $p=\diag(p_1,\dots, p_n)$
this gives
\be
\cH_2^\red(Q,p,\phi)  = \frac{1}{2} \sum_{i=1}^n p_i^2 + \frac{1}{8} \sum_{j\neq l}
\frac{\vert \phi_{jl}\vert^2}{\sin^2\frac{q_j - q_l}{2}},
\label{S23}\ee
which is a standard spin Sutherland Hamiltonian.
The last term in the Poisson bracket (\ref{S11}) represents the Poisson space $\u(n)^*//_0\T^n$,
and only gauge invariant
functions of the  spin variable $\phi$ appear in the model.

\bigskip
\noindent
{\bf Acknowledgements.}
This research was performed  in the framework of the project
GINOP-2.3.2-15-2016-00036 co-financed by the European Regional
Development Fund and the budget of Hungary.


\begin{thebibliography}{99}

\bibitem{AO}
G. Arutyunov and E. Olivucci,
{\it Hyperbolic spin Ruijsenaars--Schneider model from Poisson reduction},
{\tt  arXiv:1906.02619}

 \bibitem{CF}
O. Chalykh and M. Fairon,
{\it On the Hamiltonian formulation of the trigonometric spin Ruijsenaars--Schneider system},
{\tt arXiv:1811.08727}

\bibitem{F1}
L. Feh\'er, {\it Poisson--Lie analogues of spin Sutherland models},
  Nucl. Phys. B {\bf 949} (2019); {\tt arXiv:1809.01529 [math-ph]}

\bibitem{F2}
L. Feh\'er,
{\it Reduction of a bi-Hamiltonian hierarchy on $T^*\UN(n)$ to spin Ruijsenaars--Sutherland models},
arXiv:1908.02467 [math-ph], Lett. Math. Phys. {\bf 110} (2020) 1057-1079;  {\tt arXiv:1908.02467 [math-ph]}

\bibitem{FK1}
L. Feh\'er and C. Klim\v c\'\i k,
{\it Poisson-Lie generalization of the Kazhdan--Kostant--Sternberg reduction},
Lett. Math. Phys. {\bf 87} (2009) 125-138; {\tt arXiv:0809.1509 [math-ph]}

\bibitem{FP}
L. Feh\'er and B.G. Pusztai,
{\it Spin Calogero models obtained from dynamical r-matrices and geodesic motion},
Nucl. Phys. B {\bf 734} (2006) 304-325; {\tt arXiv:math-ph/0507062}

\bibitem{LX}
L.-C. Li and P. Xu,
{\it A class of integrable spin Calogero--Moser systems},
Commun. Math. Phys. {\bf 231} (2002) 257-286;
{\tt arXiv:math/0105162 [math.QA]}

\bibitem{Ma}
F. Magri, {\it A simple model of the integrable Hamiltonian equation},
J. Math. Phys.  {\bf 19}, 1156-1162 (1978)

\bibitem{Per}
A.M. Perelomov, Integrable Systems of Classical Mechanics and Lie Algebras, Birkh\"auser,1990

\bibitem{RSTS}
A.G. Reyman and M.A. Semenov-Tian-Shansky,
{\it Group theoretical methods in the theory of finite-dimensional integrable systems},
pp. 116-225 in: Dynamical Systems VII, V.I. Arnold and S.P. Novikov (Eds), Springer, 1994


\bibitem{RS}
S.N.M. Ruijsenaars and H. Schneider,
{\it A new class of integrable systems and its relation to solitons},
Ann. Phys. {\bf 170} (1986) 370-405

\bibitem{STS}
M.A. Semenov-Tian-Shansky,
{\it Dressing transformations and Poisson group actions},
Publ. RIMS {\bf 21} (1985) 1237-1260

\bibitem{Sm}
R.G. Smirnov,
{\it Bi-Hamiltonian formalism: A constructive approach},
Lett. Math. Phys. {\bf 41} (1997)  333-347

\bibitem{Z}
S. Zakrzewski,
{\it Free motion on the Poisson SU(N) group},
J. Phys. A: Math. Gen. {\bf 30} (1997) 6535-6543;
{\tt arXiv:dg-ga/9612008}

\end{thebibliography}
\end{document}